\documentclass[12pt,preprint]{aastex}

\usepackage{caption}
\usepackage{graphics,epsf}
\usepackage{amsmath}                % American Mathematical Society package
\usepackage{amsfonts}               % American Mathematical Society fonts
\usepackage{amssymb}                % American Mathematical Society symbol
\usepackage{epsfig}                 % EPS figures
\usepackage{subfig}
\def \cm{~\rm{cm}}
\def \s{~\rm{s}}
\def \km{~\rm{km}}

\def \K{~\rm{K}}
\def \g{~\rm{g}}

\def \erg{~\rm{erg}}

\begin{document}

\title{NUCLEOSYNTHESIS OF R-PROCESS ELEMENTS BY JITTERING JETS  IN CORE-COLLAPSE SUPERNOVAE }

\author{Oded Papish\altaffilmark{1} and Noam Soker\altaffilmark{1}}

\altaffiltext{1}{Department of Physics, Technion -- Israel Institute of Technology, Haifa
32000, Israel;  papish@physics.technion.ac.il; soker@physics.technion.ac.il}

\begin{abstract}
We calculate the nucleosynthesis inside the hot bubble formed in the
jittering-jets model for
core collapse supernovae (CCSNe) explosions, and find the formation of
several$\times 10^{-4} M_\odot$ of r-process elements.
In the jittering-jets model fast jets launched from a stochastic accretion
disk
around the newly formed neutron star are shocked at several thousands km,
and form
hot high-pressure bubbles. These bubbles merge to form a large bubble that
explode the star.
In the current study we assume a spherically symmetric homogenous bubble,
and follow its evolution
for about one second during which nuclear reactions take place.
The jets last for about one second, their velocity is $v_j=0.5c$, and their total energy is $\sim 10^{51} \erg$.
We use jets' neutron enrichment independent on time, and follow the nuclear reactions to the formation of the seed nuclei up to $Z \leq 50$,
on which more neutrons will be
absorbed to form the r-process elements.
Based on the mass of the seed nuclei we find the r-process element mass in
our idealized model to be
several$\times 10^{-4} M_\odot$, which is slightly larger than the value deduced from
observations. More realistic calculations that relax the assumptions of a homogenous bubble and constant jets composition might lead to agreement with observations.
\end{abstract}

% ==========================================================
\section{INTRODUCTION}
\label{sec:intro}
% ==========================================================

The mechanism for the explosion of core-collapse (CC) supernovae (SNe) is still unknown.
Most popular are models based on explosion driven by neutrinos.
Less popular are models based on jet-driven explosions.
(e.g. \citealt{LeBlanc1970,  Meier1976, Bisnovatyi1976, Khokhlov1999,
MacFadyen2001, Hoflich2001, Fargion2003, Woosley2005, Couch2009,Couch2011,Lazzati2011}).
Recent observations (e.g., \citealt{Wang2001, Leonard2001, Leonard2006, Elmhamdi2003, Chugai2005, Smith2011})
that found asymmetry in CCSNe suggest that jets might indeed play an important role in at least some CCSNe.

In neutrino-driven models where jets play no role at all (e.g. \citealt{Bethe1990, Nordhaus2010, Brandt2011, Hanke2011}), the neutrinos are absorbed near
the stalled-shock at $\sim 300 \km$, and revive  the shock. Namely, they
eject the material in that region.
In some theoretical studied the jets were injected at large
distances beyond the stalled-shock radius
(e.g., \citealt{Khokhlov1999, Hoflich2001, Maeda2003, Couch2009,Couch2011}).
\citealt{Hoflich2001} injected slow and fast jets at a radius of $1200 \km$ in a helium star for about two seconds.
They show the possibility of exploding the star with slow  jets. Their results show that the polarization in SN199em is consistent with slow jets.
In model m2r1hot of \citealt{Couch2011} a jet was injected near the speed of sound, leading to the formation of a hot bubble.
In others, like \citet{MacFadyen2001}, the jets were injected much closer to
the neutron star (NS), at $50 \km$.
In the simulations of  \citet{MacFadyen2001} the jets were injected at
a much later time in the explosion, and are less relevant to our goal of
exploding a star with jets.
\citet{kohri2005} propose that disk-wind energy is able to revive a stalled
shock and
help to produce a successful supernova explosion.

Our \textit{jittering-jet model} for explosion (Soker 2010; Papish \& Soker 2011, hereafter Paper 1 ) is based on the
following points, that differ in several
ingredients from the models cited above (for more detail see
Paper 1).
(1) We don't try to revive the stalled shock. To the contrary. Our model
requires the material near the stalled-shock
to fall inward and form an accretion disk around the newly born NS or black
hole (BH).
(2) We conjecture that due to stochastic processes and the stationary
accretion shock instability
(SASI; e.g. \citealt{Blondin2007}) segments of the post-shock accreted gas
(inward to the stalled shock wave) possess
local angular momentum.
When they accreted they form and accretion disk with rapidly varying axis
direction.
(3) We assume that the accretion disk launches two opposite jets. Due to the
rapid change in the disk's axis, the
jets can be intermittent and their direction rapidly varying. These are
termed jittering jets.
(4) We show in Paper 1 that the jets penetrate the infalling gas
up to a distance of few$\times 1000 \km$, i.e.,
beyond the stalled-shock. However, beyond few$\times 1000 \km$ the jets
cannot penetrate the gas any more because of their
jittering. The jittering jets don't have the time to drill a hole through the ambient gas before their direction changes;
they are shocked before penetrating through the ambient gas.
This condition can be met if the jets' axis rapidly changes its direction.
This process of depositing jets' energy into the ambient medium to prevent further accretion is termed
the
{\it penetrating jet feedback mechanism.}
(5) The jets deposit their energy inside the star via shock waves, and form
two hot bubbles, that eventually merge and
accelerate the rest of the star and lead to the explosion. In section
\ref{sec:self} below we use self similar calculations
to further explore this process.
(6) The jets are launched only in the last phase of accretion onto the NS.
For the required energy the jets must be launched
from the very inner region of the accretion disk.

Nucleosynthesis can occurs in the expanding jets (or disk winds) and in
the postshock region.
Previous studies include \citet{Cameron2001} who discussed the
nucleosynthesis inside jets launched at a velocity of
$0.5 c$ from an accretion disk around a rapidly rotating proto NS.
He suggested the possibility of creating r-process elements inside the jets.
\citet{Nishimura2006} simulated the r-process nucleosynthesis during a jet
powered explosion. In their simulation a rotating star with a magnetic field induces a jetlike outflow during the collapse which explode the star.
Neutrinos play no role in the simulation.
Unlike their model, our model does not have a large scale rotation of the star, and the jets penetrate farther away creating hot bubbles.

In our jittering-jets explosion model the jets are launched close to the NS where the
gas is neutron-rich (e.g., \citealt{kohri2005}).
In section \ref{sec:nuclear} we examine the implications of this on the
nuclear reactions (nucleosynthesis)
in the inflated hot bubble. The properties of the bubbles are as derived in
section \ref{sec:self}.
Our discussion and summary are in section \ref{sec:summary}.

% ==========================================================
\section{DYNAMICAL EVOLUTION OF THE INFLATED BUBBLES}
\label{sec:self}
% ==========================================================

In this section we describe an approximate model for the inflated bubbles and their dynamical evolution.
The analysis here is similar to that conducted by \citet{Volk1985} to study the evolution of the spherical hot bubble in planetary nebulae.
During the active phase of the jets we derive a self-similar analytical solution to the gas-dynamical equations.
At later times the solution is numerical.

\begin{figure}[!]
  \centering
    \includegraphics[width=0.7\textwidth]{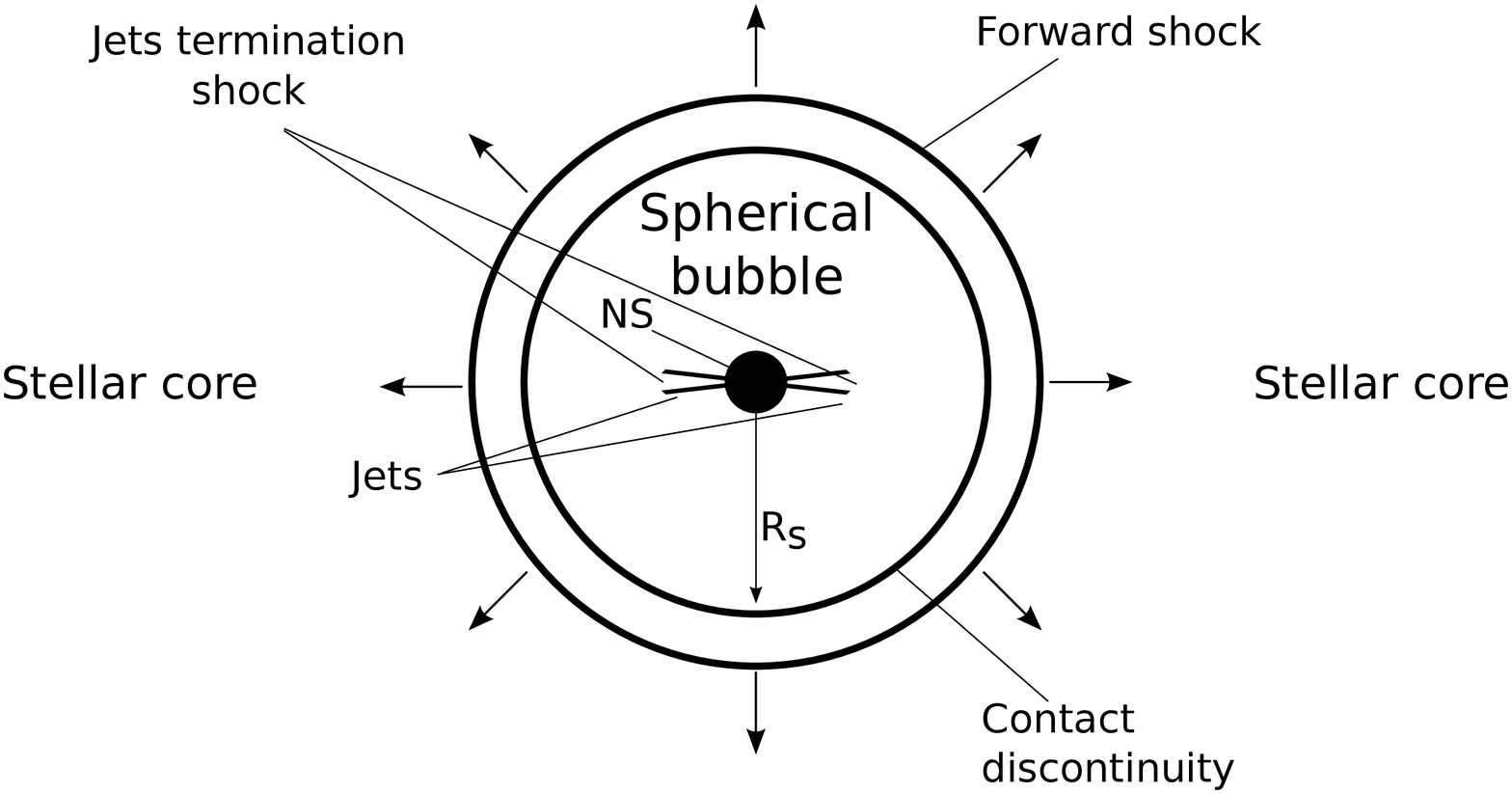}
      \caption{Schematic drawing of the inflated spherical bubble. The spherical bubble is powered by jittering jets, i.e.,
      they change their direction at a high rate,  launched from an accretion disk around the newly formed neutron star.
This spherical bubble explodes the star according to our model.
R-process elements are fused inside the bubble. The typical radius during the jets' active phase is 3000-10000 km.}
      \label{fig:bubble}
\end{figure}

The jittering jets form wide bubbles that occupy most of the volume up to the
distance they have reached; eventually the bubbles merge  (see Paper 1).
Our basic assumption is therefore that the two inflated bubbles merge to form one large bubble.
The low density high energy volume inside $R_s$  is termed
hereafter the `spherical bubble'.
If we equate the volume of the assumed spherical bubble with the total volume of the wide inflated bubbles,
the radius of the spherical bubble $R_s$ is only slightly smaller that the distance the inflated bubbles have reached.
This assumption leads to a spherically symmetric flow that allows a self similar solution for constant power jets.
We also assume that the gas inside the bubble is homogeneous, i.e.,
the composition, density, and pressure are constant inside the bubble.
The energy of the jets injected into the bubble with a power of $\dot E_j$ forces the bubble to expand.
The expanding spherical bubble pushes the dense surrounding gas supersonically outward, forming a forward
shock ahead of this dense shell. The boundary of the dense shell and the bubble is a contact discontinuity,
across which the pressure is constant but not the density or the composition. This flow structure is drawn
schematically in Fig. \ref{fig:bubble}.  The mass $M_s$ in the shell is the swept-up ambient gas.
As the dense shell is thin, we take the radius of the forward shock to be equal to the radius of the
contact discontinuity (which is the radius of the spherical bubble) $R_s$.
The pre-shock (up-stream) ambient density profile at radius $r>R_s$, is taken to be a power law,
with the scaling from \citet{wilson1986} and \citet{Mukami2008} (see Paper 1)
\begin{equation}
\rho_s(r) =  A r^{\beta} =
 1.3 \times 10^{10} \left( \frac {r}{100 \km} \right)^{-2.7} \g \cm^{-3}, \quad 30 \la r \la 10^4 \km
\label{eq:rhos}.
\end{equation}
% FFFFFFFFFFFFFFFFFFFFFFFFFFFFFFFFFFFFFFFFFFFFFFFF
% FFFFFFFFFFFFFFFFFFFFFFFFFFFFFFFFFFFFFFFFFFFFFFFF

The spherical flow obeys the following conservation equations \citep{Volk1985}
\begin{equation}
\frac{dM_s}{dt}=4 \pi R_s^2 \rho(R_s) \dot R_s,
\label{eq:mass}
\end{equation}
\begin{equation}
\frac{d}{dt} \left( M_s \dot R_s \right ) = 4 \pi R_s^2 P \dot R_s,
\label{eq:momentum}
\end{equation}
\begin{equation}
\frac{d}{dt} \left ( 4 \pi R_s^3 P \right ) = \dot E_j - 4 \pi R_s^2 P \dot R_s,
\label{eq:energy}
\end{equation}
where $P$ is the pressure inside the bubble.
Equations (\ref{eq:mass}) - (\ref{eq:energy}) describe the conservation of mass, momentum, and energy respectively.
The energy inside the bubble includes the thermal energy of the gas and the radiation energy. We neglect losses by neutrinos
(see Paper 1) and energy production and sink from nuclear reactions.

During the active time period of the jets the solution to equations (\ref{eq:mass}) - (\ref{eq:energy}) is a self-similar solution,
which we take in the form
\begin{equation}
R_s(t) = R_0 t^\alpha.
\end{equation}
Using equations (\ref{eq:mass}) - (\ref{eq:energy}) we get the following parameters
\begin{equation}
\alpha = \frac{3}{\beta+5},\quad R_0^{\beta+5} =  \frac{(\beta+3)(\beta+5)^3}{12 \pi A (2 \beta + 7)(\beta+8)}\dot E_j .
\end{equation}
A short time of $\sim 0.1 \s$ after the jets launching process ceases, energy injection to the bubble ends. The time delay comes from the jets'
crossing time from the NS to the bubble. At that moment the self-similar solution no longer holds, and we need to turn to a numerical solution.

\begin{figure}[h!]
\includegraphics[width=1.0\textwidth]{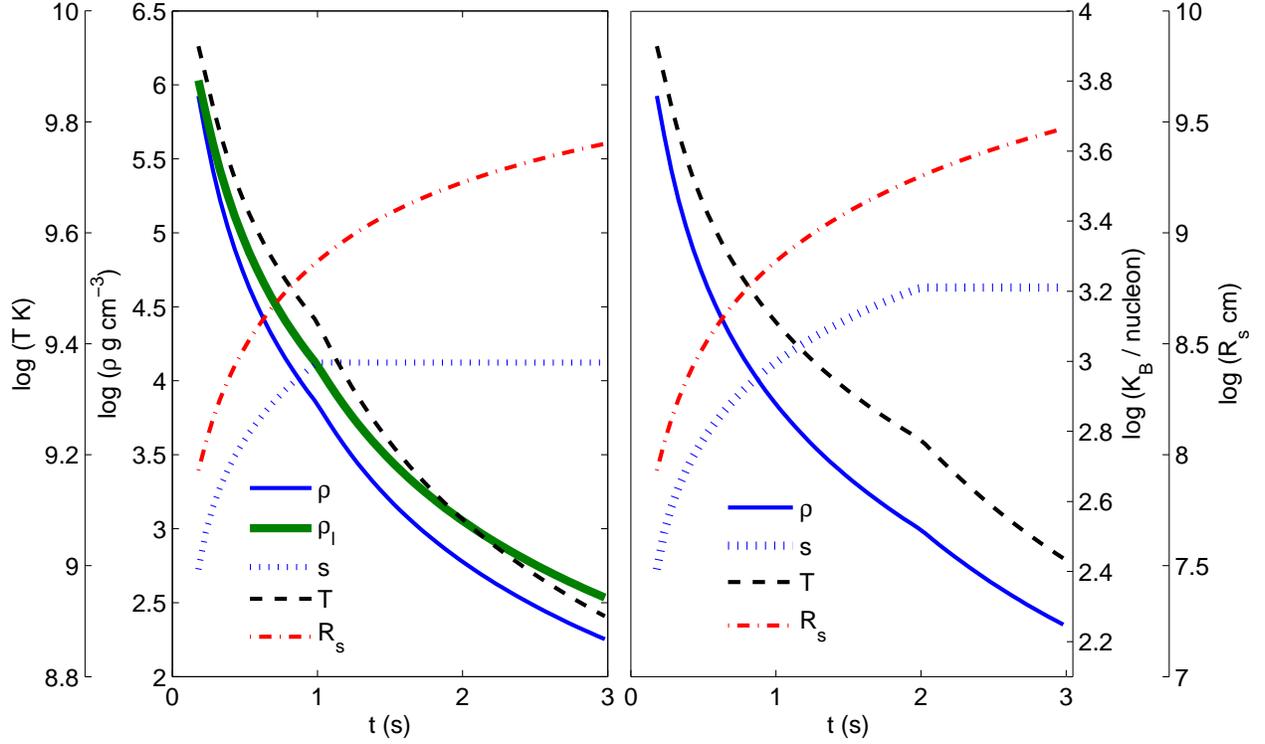}
\caption{Left: Radius $R_s$, Temperature $T$, density $\rho$, and
entropy $s$ of the spherical bubble as a function of time for the
model with jets' power of $\dot E=1.7 \times 10^{51} \erg \s^{-1}$
and a jets' active phase lasting $t_s=1 \s$.
$\rho_l$ is the density for a case with $\dot E_j=1 \times 10^{51} \erg \s^{-1}$
for the same duration of $T_s= 1 \s$.
Right: the same but for a model with $\dot E=0.85 \times 10^{51} \erg
\s^{-1}$ and an active phase duration of $t_s = 2 \s$ }
\label{fig:bubbles}
\end{figure}

The jets' are assumed to be injected at $r \sim 15 \km$ for a time period of $t_s=1-2 \s$. Here we take the velocity of the jets to be $150,000 \km \s^{-1}$ \citep{Cameron2001}.
This velocity is larger than the velocity used in Paper 1 as it better fits the escape velocity from the surface of the newly formed NS.
The velocity is chosen to be the escape velocity from the neutron star, as we are interested in the jets emerging from
the neutron star vicinity. Some studies inject the jets at distances of $>1000 \km$ at lower velocities,
e.g \citet{Couch2011}, while others use relativistic jets, e.g., \cite{Lazzati2011}, who inject the jets at $10^4 \km$ within several seconds. 
The total mass carried by the two jets is either $0.006 M_\odot$ or $0.0036 M_\odot$, corresponding to a total injected
energy of $E=1.7 \times 10^{51} \erg$ or $E=1.0 \times 10^{51} \erg$ respectively.
For the parameters of $\dot E=1.7 \times 10^{51} \erg \s^{-1}$ , and active phase time of $t_s = 1 \s$,
for example, the solution during the jets' active phase is
\begin{equation}
R_s(t) = 6.6\times 10^8\: t^{1.3} \cm , \quad 0 < t < t_s=1 \s.
\end{equation}
For later times we numerically integrate equations (\ref{eq:mass}) - (\ref{eq:energy}) with $\dot E_j=0$, and using the the results of the self similar solution at $t = t_s$  as initial conditions. For a check, we also numerically integrate the equations for the full time of the solution. The numerical solution coincides with the self-similar solution after a very short time. The full numerical solution for three cases are shown in Fig. \ref{fig:bubbles}. The plot shows the radius $R_s$, temperature $T$, density $\rho$, and entropy $s$ of the bubble as a function of time. The parameters for the three cases are given in the figure caption. The jets' power was taken to be a constant for $0 < t < t_s$, and $\dot E = 0$ for later times. This gives the little bump at $t=t_s$ in the graph.

The two cases differ by jets' power and having active time of $t_s=1 \s$, are  similar
in their general behavior.
As well, changing the active phase duration from $t_s=1 \s$ to $t_s=2 \s$ does not make large differences in the dynamical properties (the extra density line in the left panel of fig. \ref{fig:bubbles}.). All cases lead to explosion.
Later we will show that these cases are different in nucleosynthesis outcomes.

% FFFFFFFFFFFFFFFFFFFFFFFFFFFFFFFFFFFFFFFFFFFFFFFF
% FFFFFFFFFFFFFFFFFFFFFFFFFFFFFFFFFFFFFFFFFFFFFFFF

For the typical parameters expected in the model the main results of this section are as follows.
(1) The spherical bubble reaches a typical radius of $\sim 10^4 \km$ at the end of the jets' active phase.
(2) The temperature relevant for nucleosynthesis ($T\sim 3 \times 10^9 \K$) occurs at about one seconds from the beginning of the jet injection.
(3) The density in the bubble of $\sim 10^4 \g \cm^{-3}$ at that time implies that nuclear reactions will be of a high enough rate to be significant.
For that, in the next section we study the nucleosynthesis inside the bubble. \\

% ==========================================================
\section{NUCLEOSYNTHESIS INSIDE THE BUBBLE}
\label{sec:nuclear}
% ==========================================================

In the previous section we found the temperature inside the bubble to start at $\sim 10^{10} \K$, and to decrease due to adiabatic cooling to $2.5 \times 10^9 \K$
in $\sim 1 \s$.  The relevant nuclear reactions inside the hot bubble start when the temperature is about $T \simeq 9 \times 10^9 \K$ and stop at $T \simeq 2.5 \times 10^9 \K$. During this time fresh material is injected into the bubble from the jets.
As the jets are launched from very close to the neutron star, they are composed of highly enriched neutron material \citep{kohri2005}. During the expansion of the jets they adiabatically cool, and nucleons might fuse to give $\alpha$ particles and heavier nuclei (e.g., \citealt{Cameron2001, Maeda2003, Fujimoto2008}).
However, the jets are eventually shocked with a post shock temperature of $>10^{10} \K$. At that temperature all nuclei rapidly disintegrate, and a gas composed of free nucleons is formed.
Our nucleosynthesis calculations start from the free-nucleons post shock jets' gas.
At $t \simeq 0.2 \s$ when the temperature inside the spherical bubble has dropped to $T \simeq 9 \times 10^9 \K$, the free nucleons fusion rate overcomes the disintegration rate and $\alpha$ particles start to be accumulated.
During the time up to $\sim 1 \s$ the temperature drops further and $\alpha$ particle fuse to form heavier nuclei until $\alpha$ freeze-out is reached.

The nuclear reaction network is similar to that given in \citet{woosley1992}. The reaction rates are taken from the JINA Reaclib Database \citep{Cyburt2010}, and include reactions with 1,2, and 3 body interactions and beta decays. The reaction network is integrated assuming a uniform composition in the bubble and a continues injection of protons and neutrons from the jets until the jets terminated. For the electron fraction $Y_e$ we use values for the accretion disk around a neutron star as calculated by  \citet{kohri2005}.
For parameters relevant to our model we find from the calculations of \citet{kohri2005} that the neutron to proton ratio is in the range  $n/p \simeq 5-10$, namely, $Y_e \simeq 0.09-0.17$.   Here we integrate the reaction network for 3 different electron fractions $Y_e = 0.09, 0.17, 0.25$. This correspond to a neutron to proton ratio of $n/p = 10, 5 ,$ and $3$.
We study nucleosynthesis for jets' active phase of $1 \s$ and $2 \s$.
The network is solved independently of the hydrodynamics solution of section \ref{sec:self} \citep{hix1999}.

\begin{figure}[h!]
\includegraphics[width=0.9\textwidth]{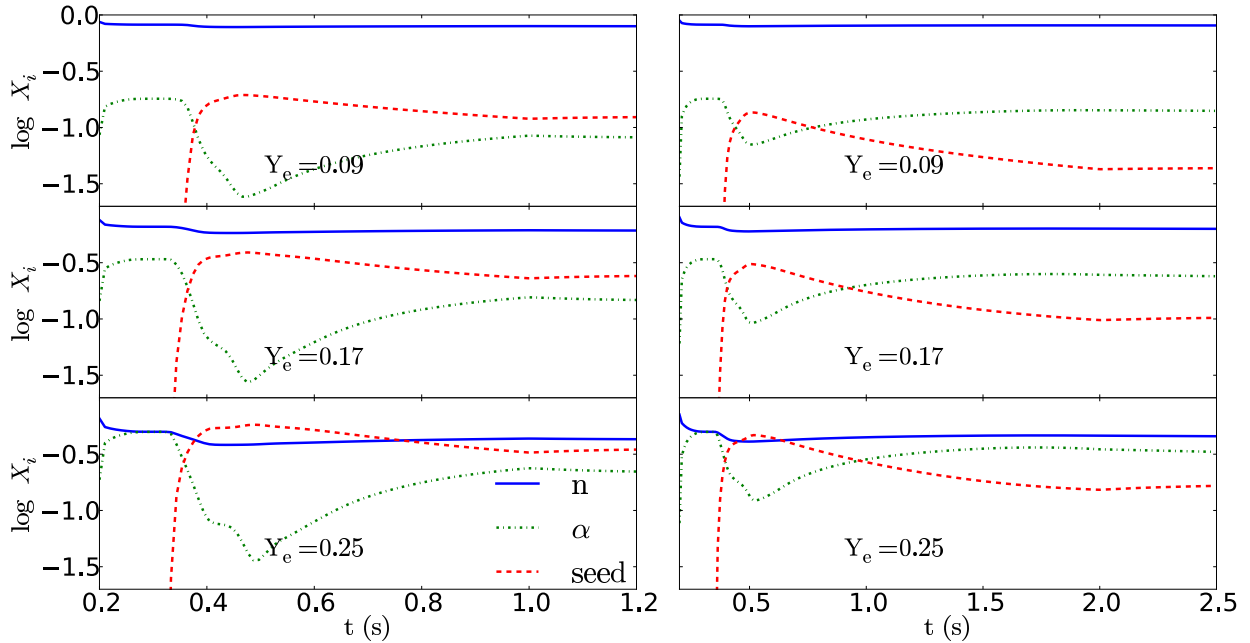}
\caption{
Evolution of the mass fraction of neutrons, $\alpha$ particles, and seed nuclei for the r-process.
Left: Model I with $t_s = 1\s$ and power of $\dot E = 1.7 \times 10^{51} \erg$. Right: Model III with $t_s = 2 \s$ and
power of $\dot E = 0.85 \times 10^{51} \erg$. In each panel the initial ratio of electron to nucleon $Y_e$ is given. The total mass
is $0.006 M_\odot$ for both cases.
}
\label{fig:rates}
\end{figure}

The nucleosynthesis results are summarized in Fig. \ref{fig:rates}.
The plots show the mass fraction of neutrons ,$\alpha$ particles, and of total seed elements during the evolution of the bubble. By seed elements we refer to nuclei on which further neutron capture will occur to synthesis the r-process nuclei (\citealt{woosley1992, Witti1994a}).
The inclusion of all nuclear processes beyond the seed nuclei is beyond the scope of the present paper. They will be studied in a forthcoming paper where a full multi-dimensional gasdynamical code will be used to study the interaction of the jets with the core material.

We assume that the post-shock freshly injected jets' material is fully mixed inside the bubble.  This is based on the expected formation of vortices inside the bubble by the jittering jets.
The post-shock velocity for $\gamma=4/3$ is $\sim 20,000 \km \s^{-1}$, and for a bubble's radius of $R_s \simeq  7000 \km$ the mixing time
is $\sim 0.3 \s$. This shows that the assumption is reasonable, but that mixing is not complete.
The assumption of full mixing will be relaxed in the future with full multi-dimensional numerical simulations.
Since the post-shock entropy increases as the jets are shocked at larger distances (because the density is lower), the entropy inside the bubble increase as long as we inject fresh jets' material.
At the relevant time of nucleosynthesis the entropy per nucleon reaches values of $> 100 ~\rm{K_B}/{\rm nucleon}$ as is required for r-process elements production \citep{Hoffman1997}.
Our flow structure differs from calculations where there is no mixing and the entropy is almost constant during nucleosynthesis  (e.g. \citealt{Witti1994a, Woosley1994, Arcones2007,Kuroda2008}).

As well, the continuous injection of nucleons as nucleosynthesis takes place reduces the final mass of seed nuclei.
This is seen by comparing the mass fraction of seed nuclei for the two
simulated cases, of $1 \s$ and $2 \s$ jets active phase duration (Table \ref{table:abundance} and Fig. \ref{fig:rates}).
The table shows also the sensitivity of the nucleosynthesis production to the value $Y_e$ assumed at the base of the jet.  \\ \\
% FFFFFFFFFFFFFFFFFFFFFFFFFFFFFFFFFFFFFFFFFFFFFFFF
% FFFFFFFFFFFFFFFFFFFFFFFFFFFFFFFFFFFFFFFFFFFFFFFF
%TTTTTTTTTTTTTTTTTTTTTTTTTTTTTTTTTTTTTTTTTTTTTTTTT
\begin{table}[h!]
\centering % used for centering table
\begin{tabular}{ l  c c c l c c c l c c c } %
\hline\hline %inserts double horizontal lines
           &        & Model I&        & &        &Model II&        & &        &Model III&  \\ [0.5ex] % inserts table
\cline{2-4} \cline{6-8} \cline{10-12}
%heading
%\hline % inserts single horizontal line
Total mass & $0.60$ & $0.60$ & $0.60$ & & $0.36$ & $0.36$ & $0.36$ & & $0.60$ & $0.60$  & $0.60$\\
Power      & $1.7$  & $1.7$  & $1.7$  & & $1.0$  & $1.0$  & $1.0$  & & $0.85$ & $0.85$  & $0.85$\\
$Y_e$      & $0.09$ & $0.17$ & $0.25$ & & $0.09$ & $0.17$ & $0.25$ & & $0.09$ & $0.17$  & $0.25$\\
\hline
$\rm n$    & 79\%   & 60\%   & 42\%   & & 79\%   & 60\%   & 42\%   & & 81\%   & 63\%    & 46\%\\
$\alpha$   & 8\%    & 14\%   & 21\%   & & 8\%    & 14\%   & 20\%   & & 14\%   & 24\%    & 33\%\\ % inserting body of the table
Seed       & 13\%   & 25\%   & 36\%   & & 13\%   & 25\%   & 37\%   & & 4\%    & 10\%    & 17\%\\ [1ex] % [1ex] adds vertical space
\hline %inserts single line
\end{tabular}
\caption{Mass fraction of neutrons, $\alpha$ particles and seed elements for different electron fractions $Y_e$. Models I and II are for jets with a duration of $t_s = 1 \s$. Model III is for jets with a duration of $t_s = 2 \s$. Total mass is in units of $10^{-2} M_\odot$. Power is in units of $10^{51} \erg$.}
\label{table:abundance} % is used to refer this table in the text
\end{table}
%TTTTTTTTTTTTTTTTTTTTTTTTTTTTTTTTTTTTTTTTTTTTTTTTT

The most likely case is for jets' active phase of $1-2$ seconds, although the jets' power
might vary with time (e.g. \citealt{Hoflich2001, Couch2009}).
The electron fraction is likely to be in the lower part of the range  $Y_e=0.09-0.17$,
corresponding to a neutron to proton ratio of $n/p = 5-10$ \citep{kohri2005}.
Also, the total energy might be slightly less that $1.7 \times 10^{51} \erg$ used here. From those values we
find the seed nuclei mass fraction to be $0.02-0.2 $, and the corresponding total seed nuclei
mass to be  $10^{-4} - 10^{-3} M_\odot$, with more likely values in the range
$10^{-4} - 3 \times 10^{-4} M_\odot$. After neutrons are absorbed (beyond the scope of this paper) and form the r-process elements, the total mass
of the r-process elements is $2-3$ times that of the seed nuclei.   This gives that the expected mass of the
r-process elements in our jittering-jets explosion model is $\sim {\rm several} \times 10^{-4} M_\odot$.

>From the solar abundance \citet{Mathews1990} deduced that the average mass of r-process material ejected in a CCSN is $\approx 10^{-4} M_\odot$.
Our simple and idealized model overproduces r-process elements, but not by much.
In future 3D numerical simulations three assumptions that have been used here will be relaxed.
These might reduce the production of r-process elements by a factor of $\sim 3-5$.
(1) Simulating precessing jets will cause deviation from sphericity. We expect that in some regions
the production of r-process elements will be less efficient.
(2) Adding the mass at the termination shocks of the jets will result in an inhomogeneous
bubble. Basically, the flow will differ by having regions where the matter cools adiabatically with no
addition of entropy (low entropy regions), and regions of high entropy where gas
is added.
This might lead to less efficient production of r-process elements in some regions.
(3) We will change the the neutron enrichment (or $Y_e$) of the jets with time. It is quite possible that at early times the
jets are less neutron enriched than at later time when the accretion disk is depleted and more mass comes from closer to the NS.

% ==========================================================
\section{SUMMARY}
\label{sec:summary}
% ==========================================================

Our main goal was to examine the nucleosynthesis inside the bubbles formed
by the jets in the
the \emph{jittering-jets model} for core collapse SN explosion (Paper 1).
The two jets are launched from an
unrelaxed accretion disk around the newly formed NS.
Because of stochastic accretion of mass and angular momentum the disk's axis
is rapidly changing and
the disk might even be intermittent.
The jets penetrate to a distance of ${\rm few} \times 1000 \km$ through the
infalling stellar core.
However, because of their changing direction they cannot penetrate beyond
that distance.
The jets are shocked and form hot low-density bubbles.
These high pressure bubbles explode the star.
This process where the non-penetrating jets prevent further accretion to the
center (a negative feedback)
is termed the {\it penetrating jet feedback mechanism.}

To facilitate a solution in the scope of the present paper we assumed that
the bubbles formed by
the two jittering jets merge to form one large spherical bubble, as shown
schematically in Fig. \ref{fig:bubble}.
The spherical solution under these assumptions is composed of two phases.
In the first one, the jets' active phase, energy and mass are injected into
the bubble at constant rates.
The second phase starts when the jets' cease, and the bubble starts to
expand adiabatically.
The gasdynamical equations in spherical symmetry were solved analytically
using a self-similar solution for the
jets' active phase, and numerically thereafter (section \ref{sec:self}).

Beyond the assumptions of jittering jets that have the power to explode the
star and the formation of a spherical bubble,
all quantitative parameters have been used before by some studies. We did not adjust
or played with any quantitative parameter in the
solutions presented here.
(1) The jets' total energy of $1-1.7 \times 10^{51} \erg$ is taken from the
energy required to explode CCSNe.
(2) The jets' velocity of $v_j=0.5 c$ comes from the escape velocity near
the NS surface. This
value for jets' velocity from NS has been used before, e.g.,
\citet{Cameron2001}.
(3) The energy and velocity determine the total mass carried by the jets.
(4) The $\sim 1-2$ seconds duration of the jets' active phase is similar to durations assumed
in other studies, e.g. \citet{Hoflich2001, Couch2009}.
(5) The ambient density profile (eq. \ref{eq:rhos}) is taken from
\citet{wilson1986} and \citet{Mukami2008}.
(6) The neutron fraction (or electron fraction $Y_e$) of the jets' material
 is taken from the calculation of \citet{kohri2005}.

The assumption of a spherical bubble that has the energy to explode the star
and the quantitative parameters
listed above determine the properties and evolution of the bubble. From these we calculated the nucleosynthesis inside the bubble.
In the limited scope of the present paper we numerically integrated a reaction network that follows
the fusion up to the seed nuclei with $Z \leq 50$.
The results of the nucleosynthesis calculations for the three studied cases
are presented in
Table \ref{table:abundance} and Fig. \ref{fig:rates}. We note that Ye = 0:25 is above
the expected value during the main phase of the jets \citep{kohri2005}, but might be applicable at
early time before the NS is fully relaxed by neutrino cooling.

During the integration of the network an $\alpha$ freeze-out is reached,
i.e., when the number of $\alpha$ particles does
not change anymore.
We take the heavy nuclei from the $\alpha$ freeze-out to be the seed
elements for r-process elements. From the mass of seed elements for the typical parameters expected in this
study, $Y_e=0.1$, energy of $10^{51} \erg$, and
active jets' duration of $t_s \simeq 1-2  \s$,  we estimate the total mass
of the fused r-process elements
to be ${\rm several} \times 10^{-4} M_\odot$.
This is a few times larger than $\approx 10^{-4} M_\odot$, the average mass of r-process elements per CCSN
deduced from observations \citep{Mathews1990}.
We note that in many observed cases the production of r-process elements is much below the average
(e.g., \citealt{Sneden2010} and references therein). It is possible, therefore, that the conditions used here
are met only in a fraction of CCSNe. For example, in some cases the value of $Y_e$ at the base of the jets is larger
than used here, namely $Y_e \ga 0.3$, where the number of neutrons is not sufficient to produce r-process elements. 

One strong assumption of the present work is that the bubble is homogeneous
in temperature and composition.
This assumption will be relaxed in a future study when the gasdynamical
equations will be solved
with a multi-dimensional numerical code. The accurate treatment of the
inflation process of the bubble will
justify the inclusion of a more extended nuclear reaction network.
Nevertheless, our results here strongly suggest that within the context of
the {\it jittering-jets model}
for core collapse SN explosions, the nucleosynthesis of the r-process
elements is a a likely possibility. This outcome strengthen the possibility that CCSNe are driven by jets.

%% ==========================================================
%\title{Acknowledgments:}
%\label{sec:Ack}
% ==========================================================
We thank an anonymous referee for helpful comments. 
This research was supported by the Asher Fund for Space Research at the
Technion and the Israel Science foundation.

% %%%%%%%%%%%%Refrences
% %%%%%%%%%%%%%%%%%%%%%%%%%%%%%%%%%%%%%%%%%%%%%%%%%%%%%%%%%%%%%%%%%%%%%%%%%%%%%%%%%%%%%

\label{lastpage}

\end{document}